\documentclass[letterpaper,twocolumn,english,english,showpacs,prl,amsmath,amssymb]{revtex4}
\usepackage[T1]{fontenc}
\usepackage[latin9]{inputenc}
\usepackage{verbatim}
\usepackage{amsmath}
\usepackage{amssymb}
\usepackage{graphicx}
\usepackage{esint}
\PassOptionsToPackage{version=3}{mhchem}
\usepackage{mhchem}

\makeatletter

\providecommand{\tabularnewline}{\\}

\@ifundefined{textcolor}{}
{%
 \definecolor{BLACK}{gray}{0}
 \definecolor{WHITE}{gray}{1}
 \definecolor{RED}{rgb}{1,0,0}
 \definecolor{GREEN}{rgb}{0,1,0}
 \definecolor{BLUE}{rgb}{0,0,1}
 \definecolor{CYAN}{cmyk}{1,0,0,0}
 \definecolor{MAGENTA}{cmyk}{0,1,0,0}
 \definecolor{YELLOW}{cmyk}{0,0,1,0}
}

\usepackage[abs]{overpic}

\usepackage{babel}

\newcommand{\E}[0]{\ensuremath{\mathcal{E}}}

\usepackage{babel}

\usepackage{babel}

\usepackage{babel}

\usepackage{babel}

\usepackage{babel}

\usepackage{babel}

\makeatother

\usepackage{babel}
\begin{document}

\title{Micromotion-induced Limit to Atom-Ion Sympathetic Cooling in Paul
Traps}

\author{Marko Cetina}

\email{marko.cetina@uibk.ac.at}

\selectlanguage{english}%

\affiliation{Department of Physics, MIT-Harvard Center for Ultracold Atoms, and
Research Laboratory of Electronics, Massachusetts Institute of Technology,
Cambridge, Massachusetts 02139, USA}

\author{Andrew T. Grier}

\affiliation{Department of Physics, MIT-Harvard Center for Ultracold Atoms, and
Research Laboratory of Electronics, Massachusetts Institute of Technology,
Cambridge, Massachusetts 02139, USA}

\author{Vladan Vuleti\'{c}}

\affiliation{Department of Physics, MIT-Harvard Center for Ultracold Atoms, and
Research Laboratory of Electronics, Massachusetts Institute of Technology,
Cambridge, Massachusetts 02139, USA}

\date{\today}

\begin{abstract}
We present and derive analytic expressions for a fundamental limit
to the sympathetic cooling of ions in radio-frequency traps using
cold atoms. The limit arises from the work done by the trap electric
field during a long-range ion-atom collision and applies even to cooling
by a zero-temperature atomic gas in a perfectly compensated trap.
We conclude that in current experimental implementations this collisional
heating prevents access to the regimes of single-partial-wave atom-ion
interaction or quantized ion motion. We determine conditions on the
atom-ion mass ratio and on the trap parameters for reaching the \textit{s}-wave
collision regime and the trap ground state. 
\end{abstract}
\maketitle
The combination of cold trapped ions and atoms \cite{Grier2009,Zipkes2010a,Zipkes2010,Denschlag11,Rellergert2011,Hall2011,Haerter2012,Ratschbacher2012}
constitutes an emerging field that offers hitherto unexplored possibilities
for the study of quantum gases. New proposed phenomena and tools include
sympathetic cooling to ultracold temperatures \cite{Smith2005,Hudson09},
charge transport in a cold atomic gas \cite{Cote2000,Cote2000_mobility},
dressed ion-atom states \cite{Cote2002,Massignan05,Goold10,Gao10},
local high-resolution probes \cite{Kollath2007,Sherkunov2009} and
ion-atom quantum gates \cite{Idziaszek2007,Doerk2010}.

In contrast to atom traps based on conservative forces, Paul traps
employ radiofrequency (RF) electric fields to create a time-averaged
secular trapping potential for the ion \cite{Leibfried2003a}. The
time-varying field can pump energy into the system if the ion's driven
motion is disturbed, e.g., by a collision with an atom \cite{Major1968}.
The kinetic behavior of an ion in a neutral buffer gas has been observed
in numerous experiments \cite{Schaaf1981,Lunney1992,Cutler1985,Herfurth2001,Kellerbauer2001,Green2007,Flatt2007,Mikosch2007}.
The ion's equilibrium energy distribution was predicted analytically
\cite{Moriwaki1992}, as well as using Monte-Carlo techniques \cite{Herfurth2001,Kellerbauer2001,Schwartz2006,DeVoe2009},
and recently a quantum mechanical analysis has been performed \cite{Nguyen2012}.
For atom-ion mass ratios below a critical value, the ion is predicted
to acquire a stationary non-thermal energy distribution with a characteristic
width set by the coolant temperature \cite{DeVoe2009}.

The RF field drives micromotion of the ion at the RF frequency. At
any position and time, the ion's velocity can be decomposed into the
micromotion velocity and the remaining velocity of the secular motion.
Consider the simple case of a sudden collision with an atom that brings
the ion to rest, a process which in a conservative trap would remove
all kinetic energy. Immediately after such a collision the ion's secular
velocity is equal and opposite to the micromotion velocity at the
time of the collision. This implies that the ion's secular motion
can be increased even in a collision that brings it momentarily to
rest. Hence for cooling by an ultracold atomic gas \cite{Zipkes2010,Denschlag11},
the energy scale of the problem is no longer set by the atoms' temperature
but by the residual RF motion of the ion, caused, e.g., by phase errors
of the rf drive or by dc electric fields which displace the trap minimum
from the rf node \cite{Zipkes2011}. Such technical imperfections
have limited sympathetic cooling so far, with the lowest inferred
ion temperature on the order of 0.5~mK \cite{Haerter2012}. Full
quantum control in these systems \cite{Idziaszek2007,Doerk2010},
on the other hand, likely requires access to the smaller temperature
scales $\hbar\omega/k_{B}\sim50\mu$K for the trap zero-point motion
and $E_{s}/k_{B}\sim50$~nK for the \textit{s}-wave collision threshold
\cite{Cote2000}.

In this Letter, we show that even with the atomic gas at zero temperature
and in a perfectly dc- and rf-compensated Paul trap, a fundamental
limit to sympathetic atom-ion cooling arises from the electric field
of the atom when polarized by the ion, or equivalently, the long-range
ion-atom interaction. As the atom approaches, it displaces the ion
from the RF node leading to micromotion, whose interruption causes
heating. A second nonconservative process arises from the non-adiabatic
motion of the ion relative to the RF field due to the long-range atom-ion
potential; during this time, the trap can do work on the ion and increase
its total energy. We find that, in realistic traps, the work done
by the RF field dominates the effect of the sudden interruption of
the ion's micromotion and leads to an equilibrium energy scale that,
for all but the lightest atoms and heaviest ions, substantially exceeds
both the \textit{s}-wave threshold $E_{s}$ and the trap vibration
energy $\hbar\omega$. Our analysis shows that current atom-ion experiments
\cite{Grier2009,Zipkes2010,Zipkes2010a,Denschlag11,Rellergert2011,Hall2011,Haerter2012,Ratschbacher2012}
will be confined to the regimes of multiple partial waves and vibration
quanta, and indicates how to choose particle masses and trap parameters
in order to achieve full quantum control in future experiments. Our
analytical results are supported by numerical calculations that furthermore
reveal that in those collisions where the RF field removes energy
from the system, the atom becomes loosely bound to the ion, leading
to multiple subsequent close-range collisions until enough energy
is absorbed from the RF field to eject the atom and heat the ion.

We consider a classical model and later confirm that this assumption
is self-consistent, i.e. that the energies obtained from the model
are consistent with a classical description. An atom of mass $m_{a}$
approaches from infinity to an ion of mass $m_{i}$ held stationary
in the center of an RF quadrupole trap. At sufficiently low collision
energies, the angular-momentum barrier will be located far from the
collision point, and, once it is passed, the collision trajectory
will be nearly a straight line. We initially assume this trajectory
to be along an eigenaxis of the RF trap, resulting in a true 1D collision
in a total potential given by $V\left(r_{i},r_{a},t\right)=e\E\left(r_{i},t\right)r_{i}/2+U(r)$,
where $e$ is the ion's charge, $r_{i}$ and $r_{a}$ are the ion
and atom locations, respectively, $r=r_{i}-r_{a}$ is the ion-atom
distance, and $\E\left(r_{i},\, t\right)=gr_{i}\cos\left(\Omega t+\phi\right)$
is the RF electric field of the ion trap at frequency $\Omega$, parameterized
by its quadrupole strength $g$. The ion-atom interaction potential
at large distances is given by $U(r)=-C_{4}/(2r^{4})$ \cite{Cote2000}
with $C_{4}$ the atom's polarizability, and modeled as a hard-core
repulsion at some small distance. Since in a three-dimensional collision
the atom can approach the ion along directions that are perpendicular
to the RF field or where the RF field vanishes, we expect our analysis
to overestimate the heating by a factor of order unity, which we address
below.

As the ion is pulled from the trap origin by the long-range interaction
$U$ with the approaching atom, the oscillating electric field $\E$
causes it to execute sinusoidal micromotion with amplitude $qr_{i}/2$,
where $q=2eg/(m_{i}\Omega^{2})$ is the unitless trap Mathieu parameter.
As long as the motion of the ion relative to the atom remains slow
relative to the average RF micromotion velocity $v_{\mu m}$, the
ion's equations of motion during one RF cycle will remain linear,
and the secular motion of the ion during each RF cycle can be described
in terms of an effective conservative secular potential $V_{s}=\frac{1}{2}m_{i}\omega^{2}r_{i}^{2}+U(r)$,
where $\omega\approx q\Omega/2^{3/2}$ is the secular frequency of
the ion within the trapping potential. Associated with $V_{s}$ are
the characteristic length scale $R=C_{4}^{1/6}/\left(m_{i}\omega^{2}\right)^{1/6}$
at which the interaction potential $U$ is equal in magnitude to the
trap harmonic potential, time scale $T=2\pi/\omega$, and energy scale
$E_{R}=\frac{1}{2}m_{i}\omega^{2}R^{2}=\frac{1}{2}(m_{i}^{2}\omega^{4}C_{4})^{1/3}$.

In collisions with light atoms, $m_{a}<m_{i}$, we expect the ion
to stay confined close to the trap origin. In this case, the ion-atom
distance $r(t)$ is governed solely by the motion of the atom in the
ion-atom potential $U$, 
\begin{equation}
r\left(t\right)\approx\left(9C_{4}/\mu\right)^{1/6}\left|t\right|^{1/3},\label{eq:FreeCollision}
\end{equation}
 where the collision occurs at time $t=0$ and $\mu=m_{i}m_{a}/(m_{i}+m_{a})$
is the reduced mass of the system. The displacement $r_{c}$ of the
ion from the center of the trap at the time of the hard-core collision
with the atom can then be estimated by integrating the effect of the
force $2C_{4}/r^{5}$ exerted by the atom on the ion trapped in its
secular potential, yielding $r_{c}\approx1.11\left(m_{a}/m_{i}\right)^{5/6}R$.
In collisions with heavy atoms ($m_{i}<m_{a}$) on the other hand,
the ion responds quickly to minimize the total secular potential energy
until, at $\left(r_{i},r_{a}\right)=\left(0.29,\,1.76\right)R$, the
deformed ion's equilibrium position becomes unstable and the light
ion quickly falls towards the atom with the collision occurring at
the ion displacement $r_{c}\approx1.76R$. In general, we may express
the collision point as $r_{c}/R=\tilde{r}_{c}\left(q,\, m_{a}/m_{i}\right)/\left(1+m_{i}/m_{a}\right)^{5/6}$
where $1<\tilde{r}_{c}<2$. 

\begin{figure}
\begin{centering}
\begin{overpic}[trim=0 0 0 -40, scale=0.60, unit=1mm]{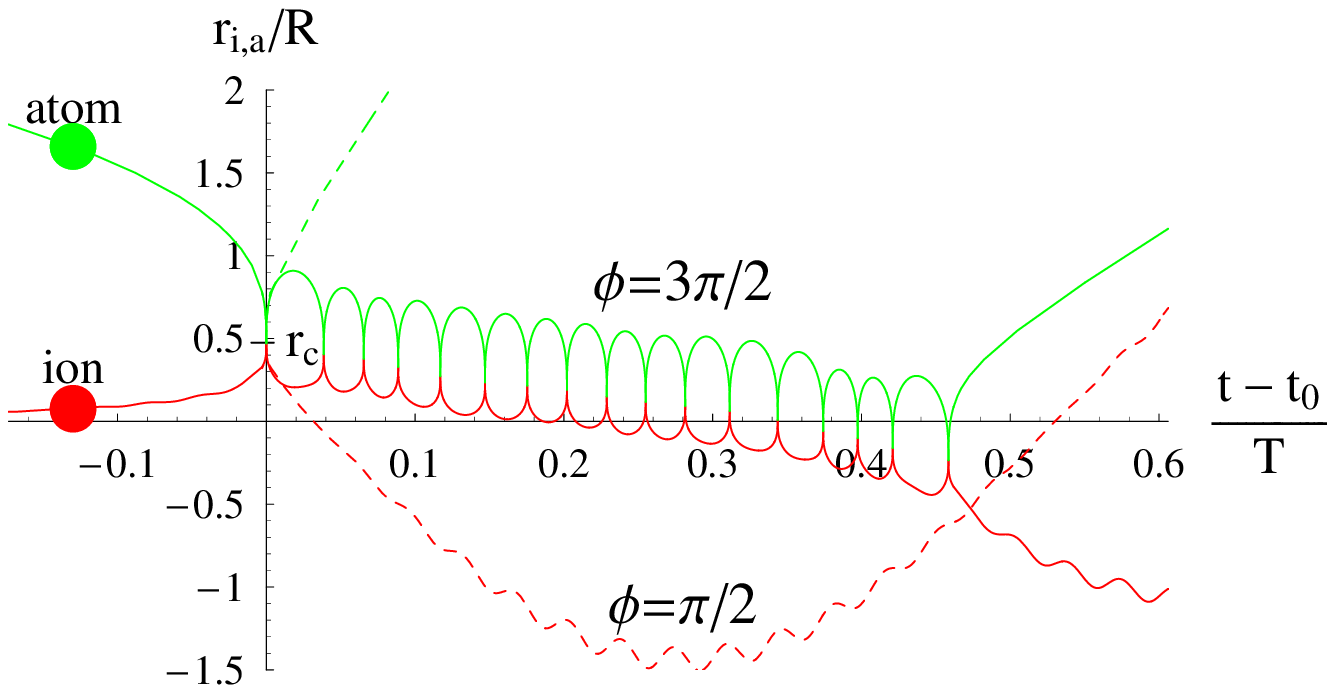}
\put(35,29){\includegraphics[scale=0.5]{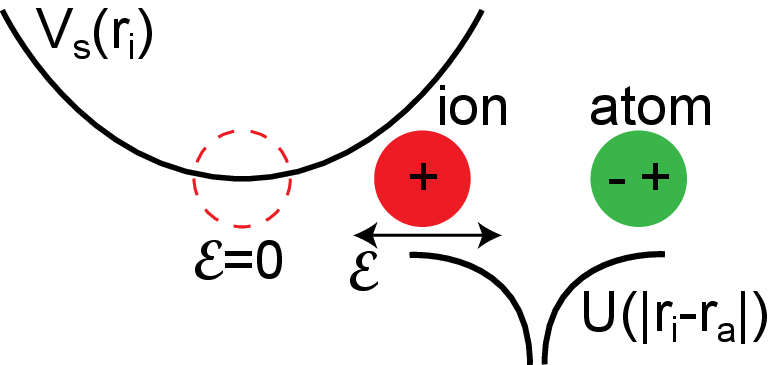}}
\end{overpic}
\par\end{centering}

\caption{\label{fig:trajs}Trajectories of an ion $r_{i}\left(t\right)$ and
an atom $r_{a}\left(t\right)$ during a classical one-dimensional
low-energy collision. The atom of mass $m_{a}$ approaches the ion
of mass $m_{i}=2m_{a}$ held in the center of a RF trap with secular
frequency $\omega=2\pi/T$ and Mathieu parameter $q=0.1$, leading
to a hard-core collision at $r_{i}=r_{a}=r_{c}$, $t=0$ and RF phase
$\phi$. For $\phi=\pi/2$ (dotted lines), the trap field adds energy
to the system, causing heating. For $\phi=3\pi/2$ (solid lines),
the RF field removes energy, binding the atom to the ion and causing
further collisions at various RF phases until enough energy is accumulated
to eject the atom.}
\end{figure}

\begin{figure}
\centering{}\includegraphics[scale=0.6]{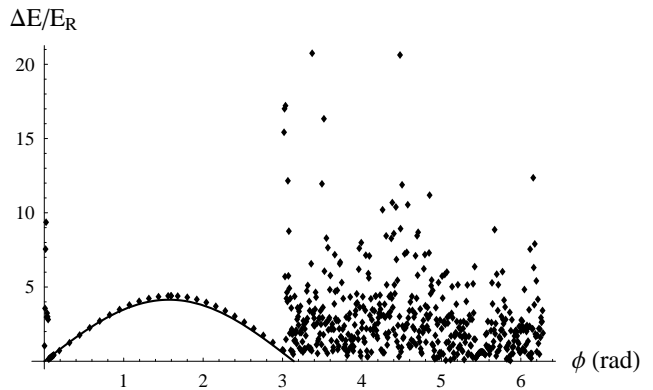}\caption{\label{fig:EvsPhi}Change $\Delta E$ in the total secular energy
of a colliding ion-atom system ($m_{a}/m_{i}=1/2$, $q=0.1$), as
a function of the RF phase $\phi$ during the first hard-core collision.
For $0<\phi<\pi$, the system undergoes only one collision with energy
gain comparable to the analytic prediction (\ref{eq:Wmax-def}) (solid
line). For $\pi<\phi<2\pi$, the atom is bound and undergoes several
collisions with the ion before eventually escaping.}
\end{figure}

Around the collision point $r_{c}$, we may expect the energy of the
system to change as long as the ion moves non-adiabatically relative
to the RF field, including the interval $-t_{1}<0<t_{1}$ around the
collision at $t=0$, during which the ion's velocity $\dot{r}_{i}$
is greater than its average micromotion velocity $v_{\mu m}\approx\omega r_{c}$.
In this regime, the trap RF field can be thought of as a time-dependent
perturbation to the ion-atom potential, doing work on the ion equal
to 
\begin{equation}
W=e\int_{-t_{1}}^{t_{1}}\E\left(r_{i}\left(t\right),\, t\right)\cdot\dot{r}_{i}\left(t\right)dt.\label{eq:RF-work-def}
\end{equation}

The change in the system's energy will depend on RF phase $\phi$
at the time of the close-range ion-atom collision, taking on the maximal
value $W_{max}$ for $\phi=\phi_{max}$. For $\dot{r}_{i}\gg v_{\mu m}$,
we may neglect the effect of the electric field on the ion's trajectory
and approximate the ion's position in terms of the free collision
trajectory $r\left(t\right)$, Eq. \ref{eq:FreeCollision}, as $r_{i}\approx r_{c}-m_{a}r/(m_{i}+m_{a})$.
In this case, the work done by the RF field can be written as $W\approx K\sin\phi$,
where 
\begin{eqnarray}
K & = & W_{0}\int_{-\Omega t_{1}}^{\Omega t_{1}}\left[\frac{\sqrt{2}\tilde{r}_{c}}{\left(3\left|\tau\right|\right)^{2/3}}-\frac{q^{1/3}}{\left(3\left|\tau\right|\right)^{1/3}}\right]\sin\left|\tau\right|d\tau\label{eq:Wmax-def}
\end{eqnarray}
and 
\begin{equation}
W_{0}=2\left(\frac{m_{a}}{m_{i}+m_{a}}\right)^{5/3}\left(\frac{m_{i}^{2}\omega^{4}C_{4}}{q^{2}}\right)^{1/3}\label{eq:W0-def}
\end{equation}
 is the characteristic scale for the work done on the ion by the RF
field. The maximal energy gain $W_{max}\approx K$ occurs for $\phi=\phi_{max}\approx\pi/2$,
corresponding to the RF field changing sign at the time of the collision.
The non-adiabatic condition $\left|\dot{r}_{i}\right|>\left|v_{\mu m}\right|$
is equivalent to $3q\Omega t<\left(2/\tilde{r}_{c}\right)^{3/2}$,
which, for the practically relevant values of $q<0.5$, will always
include the region $\left|\Omega t\right|<0.8$ where the dominant
contribution to the integral in (\ref{eq:Wmax-def}) occurs. Consequently,
we may extend the limits of integration to $\pm\infty$ to obtain
$K/W_{0}\approx1.82\tilde{r}_{c}-1.63q^{1/3}$, implying $0.7<K/W_{0}<2.8$
for $q<0.5$ and at all mass ratios. Under the same conditions, $K$
is at least three times larger than the ion's average micromotion
energy at the collision point $E_{\mu m}\approx m_{i}v_{\mu m}^{2}/2$
and the gradual energy change (\ref{eq:RF-work-def}) dominates the
effect of the sudden interruption of the ion's micromotion. Intuitively,
at low collision energies, the ion-atom potential dominates for a
longer time during which the RF field does more work on the system.
Since the trap electric field must be increased for higher RF frequencies
to preserve the ion secular potential, the heating increases with
a decrease in the Mathieu parameter ($K>15\times E_{\mu m}$ for $q<0.1$).

Since $W$ corresponds to the difference in the work done by the RF
field during the incoming and outgoing parts of the collision, the
energy change will depend on the phase $\phi$ of the RF field at
the time of the hard-core collision. For $0<\phi<\pi$, the RF field
accelerates the collision partners towards each other, increasing
the system energy; for $\pi<\phi<2\pi$, the RF field opposes the
collision, doing negative work and causing the atom to be bound to
the ion with binding energy on the order of $-W_{0}$ (Figure \ref{fig:trajs}).
Since the $r^{-4}$ potential does not possess stable orbits, bound
ion-atom trajectories will include further close-range collisions.
Depending on the RF phase during each subsequent collision, the system
will gain or lose energy on the order of $W_{0}$, leading to a random
walk in energy until the atom finally unbinds. Then $\Delta E$ depends
sensitively on the RF phases at each hard-core collision spaced in
time by many RF cycles, leading to a sharp dependence of $\Delta E$
on the RF phase $\phi$ during the initial collision for $\pi<\phi<2\pi$
(Figure \ref{fig:EvsPhi}). Using the approach below, we calculate
the net average energy gain for $0<\phi<2\pi$ and $m_{a}/m_{i}=1/2$,
$q=0.1$ below as $\left\langle \Delta E\right\rangle =1.0W_{0}$.

\begin{figure}
\begin{overpic}[trim=95 -50 0 0, scale=.62, unit=1mm]{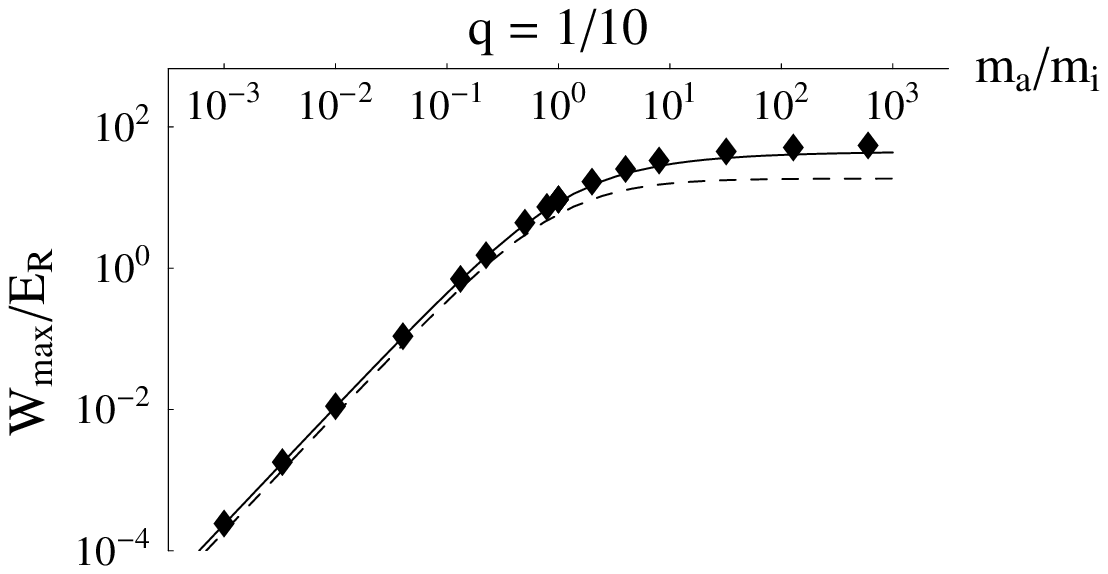}
\put(24,-15){\includegraphics[scale=0.60]{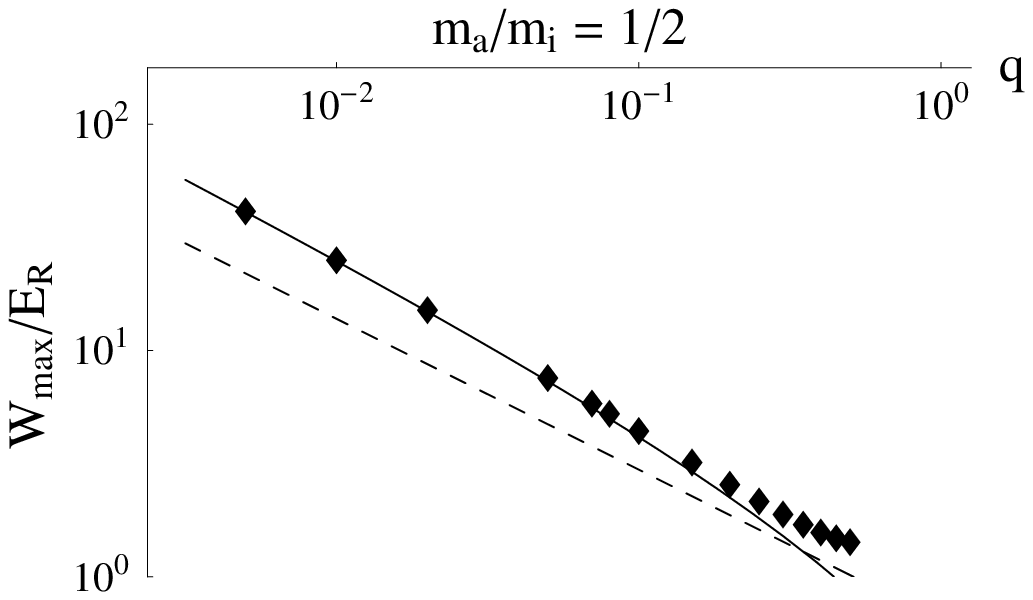}}
\put(-5,13){(a)}
\put(20,0){(b)}
\end{overpic}

\caption{\label{fig:EmaxVsma}Maximal energy gain $W_{max}$ in a low-energy
1D ion-atom collision as a function of the atom/ion mass ratio $m_{a}/m_{i}$
with Mathieu parameter $q=0.1$ (a) and as a function of $q$ with
$m_{a}/m_{i}=1/2$ (b). Data points are numeric calculations. The
solid lines correspond to $W_{max}=K$, while the dashed lines represent
$W_{max}=W_{0}$.}
\end{figure}

\begin{figure}
\begin{overpic}[trim=0 -7 0 0, scale=.60, unit=1mm]{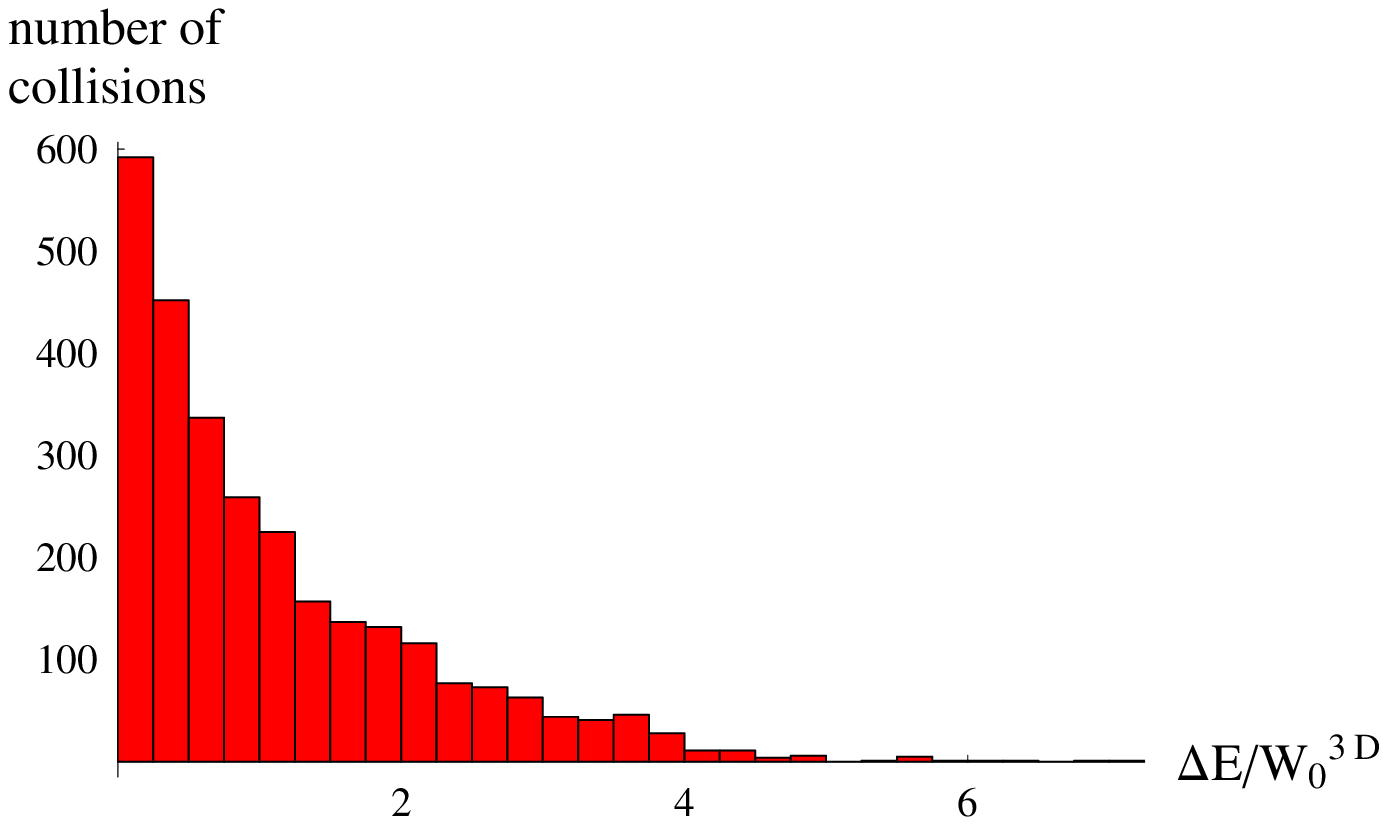}
\put(20,+11){\includegraphics[scale=0.60]{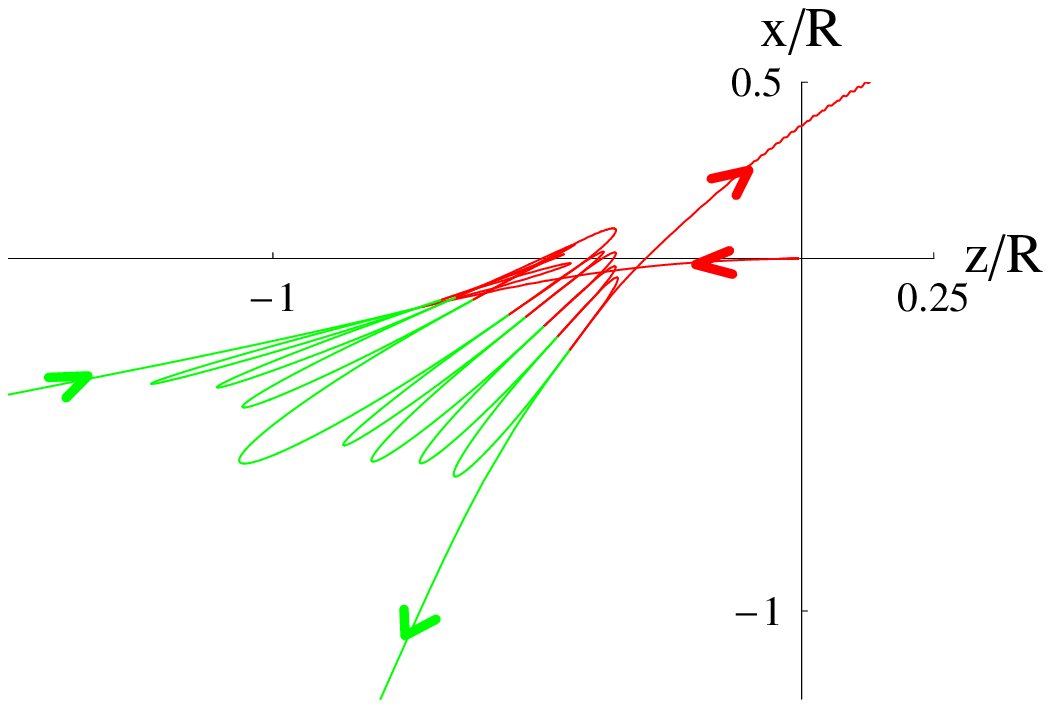}}
\put(40,45){(b)}
\put(40,-1){(a)}
\end{overpic}\caption{\label{fig:Kohl-heating}(a) Distribution of numerically computed
secular energy gains in 2883 random low-energy collisions between
a free $^{87}$Rb atom and a $^{174}$Yb$^{+}$ ion held in the three-dimensional
trap from \cite{Zipkes2010a}, in units of the 3D micromotion heating
energy scale $W_{0}^{3D}=4W_{0}/\left(3\pi\right)$. A sample ion-atom
collision trajectory is shown in the inset (b).}
\end{figure}

To quantitatively verify the above heating model, we numerically calculated
classical trajectories of low-energy 1D collisions as a function of
$\phi$, $m_{a}/m_{i}$ and $q$ (Figures \ref{fig:EvsPhi} and \ref{fig:EmaxVsma}).
The ion was initially held at the trap center while the atom followed
the analytic trajectory (\ref{eq:FreeCollision}). At a critical ion-atom
distance $r_{0}$, the ion was displaced so that the trap's secular
force would balance the atom's attraction, while keeping its velocity
zero. Choosing $r_{0}=3.8\left(1+m_{i}/m_{a}\right)^{1/3}R$ ensured
that for all the trap parameters in this paper, the ion's intial micromotion
energy was smaller than $10^{-5}W_{0}$ and both the atom's kinetic
energy and $U\left(r_{0}\right)$ were smaller than $10^{-2}W_{0}$.
The equations of motion were integrated using the Dormand-Prince explicit
Runge-Kutta method: away from collision points, the motion was integrated
as a function of time while near the collisions, the ion-atom distance
$r$ was used with a hard-core radius $\epsilon=10^{-3}\left(1+m_{i}/m_{a}\right)^{1/3}R$.
The integration was stopped when the atom reached a distance $r_{a}=2.1r_{0}$
much larger than the ion motion, at which point the total secular
energy of the system was evaluated. We confirmed the accuracy of our
integration by replacing the RF potential with a time-independent
secular potential and confirming energy conservation at the level
of $10^{-3}W_{0}$.

Figures \ref{fig:EmaxVsma}a and \ref{fig:EmaxVsma}b show the numerically
calculated maximal energy gain $W_{max}$ in the initial collision
as a function of $m_{a}/m_{i}$ and $q$. For $q\leq0.1$, the calculated
$W_{max}$ is within 30\% of the analytic prediction $W_{max}=K$;
for $q=0.5$, $m_{a}/m_{i}=1/2$, $W_{max}$ increases to $1.7K$,
partly because the analytic result does not include the micromotion
interruption%
\begin{comment}
on the order of $E_{\mu m}=0.25K$ 
\end{comment}
. The heating is insensitive to the the hard-core radius: a tenfold
increase or decrease in $\epsilon$%
\begin{comment}
or a twofold increase or decrease in $r_{0}$: WmaxVsMa\_closer\_r0.nb/WmaxVsMa\_farther\_rcoll.nb
and WmaxVsq\_closer\_rcoll.nb/WmaxVsq\_farther\_r0.nb
\end{comment}
{} changes $W_{max}$ by less than 1\%%
\begin{comment}
WmaxVsMa\_closer\_rcoll.nb/WmaxVsMa\_farther\_rcoll.nb and WmaxVsq\_closer\_rcoll.nb/WmaxVsq\_farther\_rcoll.nb
\end{comment}
. Our results are also very robust with respect to the initial conditions:
the average energy gain in Figure \ref{fig:EvsPhi} changes by less
than 5\% if the ion is started at rest in the center of the trap and
the atom at rest a distance $R$ away%
\begin{comment}
EVsPhi\_avg.nb vs. fig3new.nb
\end{comment}
.

In three-dimensional linear quadrupole RF traps, the electric field
$\E$ and the ion position $r_{i}$ in (\ref{eq:RF-work-def}) are
vectorial quantities. Assuming that the collision trajectory is still
nearly one-dimensional, averaging over the atom's approach direction
rescales the work done by the RF field, Eq. (\ref{eq:RF-work-def}),
by $4/\left(3\pi\right)$, leading to a natural 3D heating scale $W_{0}^{3D}=4W_{0}/\left(3\pi\right)$.
To check this, numerical simulations were done for the three-dimensional
quadrupole RF trap from Ref. \cite{Zipkes2010a}. We considered cold
$^{87}$Rb atoms that have passed the angular momentum barrier at
a large distance and are colliding head-on with a $^{174}$Yb$^{+}$
ion from a random direction and at a random time. The initial conditions
for numeric integration were chosen by analogy to the 1D case. Figure
\ref{fig:Kohl-heating}b shows a sample ion-atom trajectory including
multiple collisions. Due to a difference in the axial and radial frequencies
of the ion trap, the collision trajectory precesses about the $y$
axis, while remaining nearly one-dimensional close to the collision
points. A histogram of the final system energies $E$ after the atom
is ejected to infinity is shown on Figure \ref{fig:Kohl-heating}a,
with an average energy gain of $0.9W_{0}^{3D}$, confirming that the
heating is similar in one and three dimensions.

\begin{table}
\begin{centering}
\begin{tabular}{|cc|c|c|c|c|c|c|}
\hline 
ion /atom  &  & {\footnotesize $\omega/(2\pi)$} & $q$  & {\footnotesize $E_{s}/k_{B}$} & {\footnotesize $E_{DC}^{s}$} & {\footnotesize $\hbar\omega/k_{B}$} & {\footnotesize $W_{0}/k_{B}$}\tabularnewline
 &  & {\footnotesize kHz} &  & {\footnotesize nK} & {\footnotesize mV/m} & {\footnotesize $\mu$K} & {\footnotesize $\mu$K}\tabularnewline
\hline 
\hline 
$^{174}$Yb$^{+}$ $^{87}$Rb  & \cite{Zipkes2010a}  & 200  & 0.013  & 44  & 4.6  & 9.6  & 540\tabularnewline
\hline 
$^{87}$Rb$^{+}$ $^{87}$Rb  & \cite{Haerter2012}  & 350  & 0.24  & 79  & 7.7  & 17  & 210\tabularnewline
\hline 
$^{138}$Ba$^{+}$ $^{87}$Rb  & \cite{Denschlag11}  & 200  & 0.11  & 52  & 4.5  & 9.6  & 150 \tabularnewline
\hline 
$^{40}$Ca$^{+}$ $^{87}$Rb  & \cite{Hall2011}  & 110  & 0.20  & 200  & 2.6  & 5.3  & 50\tabularnewline
\hline 
$^{174}$Yb$^{+}$ $^{172}$Yb  & \cite{Grier2009}  & 67  & 0.14  & 44  & 1.6  & 3.2  & 41\tabularnewline
\hline 
$^{174}$Yb$^{+}$ $^{40}$Ca  & \cite{Rellergert2011}  & 250  & 0.25  & 270  & 14  & 12  & 32\tabularnewline
\hline 
$^{174}$Yb$^{+}$ $^{23}$Na  &  & 50  & 0.30  & 710  & 4.7  & 2.4  & 1.5\tabularnewline
\hline 
$^{174}$Yb$^{+}$ $^{7}$Li  &  & 50  & 0.30  & 6400  & 14  & 2.4  & 0.24 \tabularnewline
\hline 
\end{tabular}\\

\par\end{centering}

\caption{\label{tab:summary}The quantum $s$-wave energy limit ($E_{s}$),
the corresponding radial DC electric field $E_{DC}^{s}$ at which
the ion micromotion energy is equal to $E_{s}$, the ion trap vibrational
quantum ($\hbar\omega$), and the micromotion-induced energy scale
$W_{0}$ in various ion-atom systems. The Yb$^{+}+$Rb and Ba$^{+}+$Rb
systems have a sufficiently large ratio of elastic to inelastic collisions
to permit cooling \cite{Denschlag11,Ratschbacher2012}. In the Yb$^{+}+$Yb
and Rb$^{+}+$Rb systems, it may be possible to make charge exchange
collisions endothermic by choosing the proper isotope combination.
The Yb$^{+}+$Ca system exhibits a very large inelastic collision
rate.}
\end{table}

Table \ref{tab:summary} shows $W_{0}$ together with the $s$-wave
threshold energy $E_{s}=\hbar^{4}/(2\mu^{2}C_{4})$ and the energy
$\hbar\omega$ of a trap vibrational quantum in various experimental
systems. In the current systems \cite{Zipkes2010a,Denschlag11,Rellergert2011,Grier2009,Haerter2012},
the atom number per characteristic volume $R^{3}$ is much smaller
than one, while the atom kinetic energy is much smaller than $W_{0}$,
and we expect binary ion-atom collisions to be well described by our
model. Given the typical duration of a collision of about $T/2$ and
typical ion-atom collision rate coefficients \cite{Haerter2012},
corrections due to three-body collisions are expected at critical
atom densities on the order of $n_{c}\sim10^{14}$~cm$^{-3}$. At
these densities, three-body inelastic processes become significant,
leading to strong ion heating and loss as was recently observed \cite{Haerter2012}.
Thus best sympathetic cooling is expected in the density regime $n<n_{c}$,
where it is limited by the two-body energy scale $W_{0}$ derived
above (Eq. \ref{eq:W0-def}) to temperatures $T_{min}\sim W_{0}/k_{B}$.
In currently realized systems, $W_{0}$ is more than one order of
magnitude larger than the trap vibrational quantum and almost three
orders of magnitude larger than the $s$-wave scattering limit.

Since, for light atoms, the ratio of of heating to the $s$-wave collision
threshold scales as $W_{0}/\hbar\omega\propto\left(\omega C_{4}\right)^{4/3}m_{a}^{11/3}/m_{i}$
and the ratio of heating to the trap vibration quantum scales as $W_{0}/(\hbar\omega)\propto\left(\omega C_{4}\right)^{1/3}m_{a}^{5/3}/m_{i}$,
our model predicts that micromotion heating could be mitigated using
light atoms and heavy ions trapped in weak RF traps, limited by the
control of DC fields (since $E_{DC}^{s}\propto\omega$). In particular,
with control over the DC electric fields on the order of 10 mV/m -
an order of magnitude better than current state-of-the-art \cite{Haerter2012}
- the Yb$^{+}$/Li system may enter the $s$-wave regime. Heating
could also be decreased by employing Raman transitions to produce
molecular ions that are sufficiently bound so as not to be affected
by micromotion \cite{Nguyen2012}. Another option may be the use of
an optical trap for the ion, as was recently demonstrated \cite{Schneider2010}.

\expandafter\ifx\csname natexlab\endcsname\relax\global\long\def\natexlab#1{#1}
 \fi \expandafter\ifx\csname bibnamefont\endcsname\relax \global\long\def\bibnamefont#1{#1}
 \fi \expandafter\ifx\csname bibfnamefont\endcsname\relax \global\long\def\bibfnamefont#1{#1}
 \fi \expandafter\ifx\csname citenamefont\endcsname\relax \global\long\def\citenamefont#1{#1}
 \fi \expandafter\ifx\csname url\endcsname\relax \global\long\def\url#1{\texttt{#1}}
 \fi \expandafter\ifx\csname urlprefix\endcsname\relax\global\long\def\urlprefix{URL }
 \fi \providecommand{\bibinfo}[2]{#2} \providecommand{\eprint}[2][]{\url{#2}}

\end{document}